\documentclass[12pt]{article}
\usepackage{graphicx}
\begin{document}
%
%
%
%
\begin{center}
{\Large          General properties of the pion production 
                      reaction in nuclear matter
} 
\end{center}
%
%
\begin{center}
\normalsize
P. Camerini$^{a,b}$, E. Fragiacomo$^{a,b}$, N. Grion$^{a,}$\footnote[1]
{Corresponding author, electronic mail: Nevio.Grion@ts.INFN.it}, 
S. Piano$^{a,b}$, R. Rui$^{a,b}$, J. Clark$^c$, L. Felawka$^d$, 
E.F. Gibson$^{e}$, G. Hofman$^{d}$, E.L. Mathie$^{f}$, R. Meier$^{g}$
G. Moloney$^c$, D. Ottewell$^d$, K. Raywood$^d$, M.E. Sevior$^c$, 
G.R. Smith$^{d,}$\footnote[2] 
{Permanent address: Jefferson Lab, Newport News, VA 23006}, 
and R. Tacik$^f$.
\end{center}
%
%

\begin{center}
\small{\it 
$^a$ Istituto Nazionale di Fisica Nucleare, 34127 Trieste, Italy \\
$^b$ Dipartimento di Fisica dell'Universita' di Trieste, 34127 Trieste, 
     Italy\\
$^c$ School of Physics, University of Melbourne, Parkville, Vic., 3052,
     Australia \\
$^d$ TRIUMF, Vancouver, B.C., Canada V6T 2A3 \\
$^e$ California State University, Sacramento CA 95819, USA \\
$^f$ University of Regina, Regina, Saskatchewan, Canada S4S 0A2 \\
$^g$ Physikalisches Institut, Universit\"{a}t T\"{u}bingen, 72076 
     T\"{u}bingen, Germany. 
}
\end{center}                   
\begin{center}
\large{The CHAOS Collaboration} 
\end{center}

%
%
\setlength{\baselineskip}{2.3ex}         
{\small
The pion production reaction 
$\pi^+ \rightarrow \pi^+\pi^{\pm}$ on $^{45}Sc$ was studied at 
incident pion energies of $T_{\pi^{+}}$ = 240, 260, 280, 300, 
and 320 MeV. The experiment was performed using the $M11$ 
pion-channel at TRIUMF, and multiparticle events, 
($\pi^+,\pi^+\pi^{\pm}$) and ($\pi^+,\pi^+\pi^{\pm}p$), were detected
with the CHAOS spectrometer.  Results are reported in the form of
both differential and total cross sections, and are compared to 
theoretical predictions and the reaction phase space. The present 
investigation of the T-dependence of the 
$\pi^+ A \rightarrow \pi^+\pi^{\pm} A'$ reaction complements earlier 
examinations of the A-dependence of the reaction, which was measured 
using $^{2}H$, $^{4}He$, $^{12}C$, $^{16}O$, $^{40}Ca$, and $^{208}Pb$ 
targets at $\sim$280 MeV. Some general properties of the pion-induced 
pion production reaction in nuclear matter will be presented, based 
on the combined results of the two studies.
}

\newpage

\normalsize
%
%
{\bf 1. Introduction}

The $\pi^+\rightarrow\pi^+\pi^\pm$ reactions ($\pi 2\pi$) on
$^{45}Sc$ were studied at intermediate energies, 
$T_{\pi^+}$ = 240, 260, 280, 300, and 320 MeV.  The
A-dependence of the $\pi 2\pi$ process was studied previously on 
$^{2}H$, $^{4}He$, $^{12}C$, $^{16}O$, $^{40}Ca$ and $^{208}Pb$ 
nuclei at $T_{\pi^+}\sim$280 MeV \cite{expt:one,
expt:1.5,expt:two,expt:2.5,expt:three,expt:four,expt:five}. 
The results of those investigations, combined with the present
investigation of the $\pi 2\pi$ T-dependence, allow for the
determination of some general properties of the $\pi 2\pi$ reaction 
in nuclei, as well as those of the $\pi\pi$ system. The $^{2}H$ 
studies were performed both to understand the $\pi 2\pi$ behaviour 
on a neutron and a proton by means of the $\pi N\rightarrow\pi\pi N$ 
quasifree reaction, and to observe medium modifications of the 
$\pi\pi$ interaction through the direct comparison to the $\pi 2\pi$ 
data in nuclei. To ensure a reliable comparison, all the $\pi 2\pi$ 
data were taken under the same kinematical conditions. 

The detection of $\pi\pi$ pairs in the I=0 J=0 channel is a 
way of examining the existence of the light scalar-isoscalar 
meson, the $\sigma$ meson. The direct observation of $\sigma$'s 
in vacuum has always been frustrated by the elusiveness of the 
meson, perhaps due to the large width of its spectral function 
\cite{expt:5.5}. The threshold behaviour of the $(\pi\pi)_{I=J=0}$ 
system has been the subject of experimental investigations: the 
invariant mass intensities display a remarkable strength in nuclei, 
while they appear depleted in the nucleon (i.e., in vacuum) 
\cite{expt:five}. An interpretation of this observation relies 
on the partial restoration 
of chiral symmetry in nuclear matter. In a chiral symmetric space, 
chiral partners would have same spin but opposite parity, and be 
degenerate in mass. In this framework, the chiral partner of 
the $\sigma$ ($J^P=0^+$) is the pion ($J^P=0^-$). These mesons, 
however, have the masses which are a few hundred MeV apart. The 
shift of the $\pi\pi$ invariant mass toward the lower $2m_\pi$ in 
nuclei is taken as a signature of partial restoration of the broken 
symmetry \cite{sigma:one,sigma:two}. It is worthwhile noting that 
this process is observed in nuclei, i.e. at nuclear densities 
$\rho < \rho_n$ the saturation density. In the same 
environment, another process contributes to the reshaping of 
the $(\pi\pi)_{I=J=0}$ spectral function at around the 2$m_\pi$ 
threshold (and below it), which is due to 
the P-wave coupling of pions to {\em particle-hole} ({\em p-h}) 
and {\em delta-hole} ({\em $\Delta$-h}) configurations. The 
net result is similar, although less pronounced, to that of 
partial restoration of chiral symmetry 
\cite{ph_Dh:one,ph_Dh:two,ph_Dh:three,ph_Dh:four,ph_Dh:five}. 

Correlated pion pairs have also been studied in a unitary 
chiral nonperturbative model \cite{sigma:six}. In this approach,
the $\sigma$ resonance in vacuum is dynamically generated by 
the strong $\pi\pi$ interaction. In nuclear matter, the $\sigma$ 
properties are only weakly modified \cite{sigma:seven}, whereas
the $(\pi\pi)_{I=J=0}$ interaction is altered by the 
P-wave coupling of pions to {\em p-h} and {\em $\Delta$-h} 
states. In the I=0 J=0 channel, $ImT_{\pi\pi}$ enhances its 
strength at 2$m_\pi$ as pion pairs probe increasing $\rho$'s. 
The increase is detectable at $\rho < \rho_n$. The pion 
production reaction in nuclei proceeds via the quasifree 
$\pi N \rightarrow \pi\pi N$ elementary process \cite{ph_Dh:six}, 
and the nuclear influence on the process is accounted for by a 
complete model \cite{ph_Dh:five}, which is considered the 
{\em state-of-the-art} in this field. 

The purpose of the present article is to give an account of
new results on the T-dependence of the $\pi 2\pi$ reaction, 
and to examine the data in relation to the above mentioned 
theoretical results.
In addition, complemented by the earlier studies of the
A-dependence, general properties of the $\pi 2\pi$ 
process in nuclear matter are sketched out.
The $\pi 2\pi$ data were collected 
under the same kinematical conditions to allow for a 
straightforward comparison of the results. Pion pairs were analysed 
down to opening angles $\sim 0^\circ$ in order to determine the 
$\pi\pi$ invariant mass at the $2m_{\pi}$ threshold, and the energy 
of incident pions was limited to 320 MeV to reduce the final state 
interactions of pions with the residual nucleus. The article 
is organised as follows: some features of the experiment are 
reported in Sec. 2. Sec. 3 deals with the data analysis. Exclusive
$\pi 2\pi$ data at 300 MeV are discussed in Sec. 4. The $\pi 2\pi$ 
model used for comparison is presented in Sec. 5. The T-dependence
of the reaction is reported in Sec. 6. Other existing results are 
mentioned in Sec. 7. Finally, conclusions are summarised in Sec. 8.

{\bf 2. The experiment at TRIUMF}

The aim of the experiment was the measurement of differential
cross sections for the $\pi^+  \rightarrow \pi^+\pi^{\pm}$ 
reactions at several incident pion kinetic energies. This 
required single particle mass-identification and vector momentum 
determination for each event.

2.1 The pion beam and the beam counting

\begin{table}[tbc]
\caption[Table]
{Characteristics of the particle beam: Central momentum (p, [MeV/c])
and energy ($T$, [MeV]), momentum spread ($\delta$p/p at FWHM, [\%]), 
intensity ($N$, [$10^6$particles/s]), and composition 
\vspace{0.5cm}($\pi[\%],p[\%]$).}
\begin{center}
  \begin{tabular}{cccc}  \hline \hline 
 p, \, T               & $ \delta$p/p & $N$ & $\pi,$ \, p \\ \hline 
353, 240               &   2.0        & 1.6 & 95.5, 4.5  \\ 
374, 260               &   1.1        & 3.4 & 94.3, 5.7  \\ 
396, 280               &   1.1        & 2.6 & 94.9, 5.1  \\ 
417, 300               &   1.0        & 2.4 & 95.7, 4.3  \\ 
438, 320               &   1.6        & 1.5 & 96.4, 3.6  \\ 
\hline\hline
  \end{tabular}
\end{center}
\end{table}
The experiment was performed at the TRIUMF Meson 
Facility, using positive pions from the M11 channel.
The final focus was located at the centre of the 
detector.  Pion energies were
240, 260, 280, 300, and 320 MeV.
Positive pions were accompanied by the $\pi^+$'s decay products, 
i.e. $\mu^+$'s and $e^+$'e, and $p$'s. However, the contamination 
of positrons and positive muons was found to be less than 1\% at all 
energies. Protons were initially separated from pions by means of an 
absorber placed in the mid-plane of M11, and finally intercepted by 
the slits of the channel. Protons escaping the slits were constantly 
monitored through their time-of-flight along the beam line 
($\sim$13.5 m long). The beam characteristics are summarised in 
Table 1. The target used was a solid slab of $^{45}Sc$ 0.558 
g/cm$^2$ thick. The target area exposed to the pion beam was 
4.25$\times$3.03 cm$^2$, which could contain a large fraction of 
the beam envelope.

The number of pions impinging upon the target was constantly 
monitored by an in-beam scintillator counter, which was located 
at about 0.8 m from the target. Therefore, the real number 
of pions arriving at the target was corrected for the rate 
of pion decay, which was $\sim$3.5\% at $T_{\pi^{+}}$=280 MeV. 
In addition, a rate correction due to multiple pions per beam 
burst was applied; this was, for example, $\sim$4.9\% at 
$T_{\pi^{+}}$=280 MeV. Detailed calculations are reported in 
Ref. \cite{expt:five}.  

2.2 The CHAOS spectrometer 

Multiparticle charged events were detected by the CHAOS 
spectrometer\cite{CHAOS:one}. CHAOS consists of a dipole 
magnet, four cylindrical wire chambers (WC) and a cylindrical 
\begin{figure}[h]
 \centering
  \includegraphics*[width=0.4\textwidth]{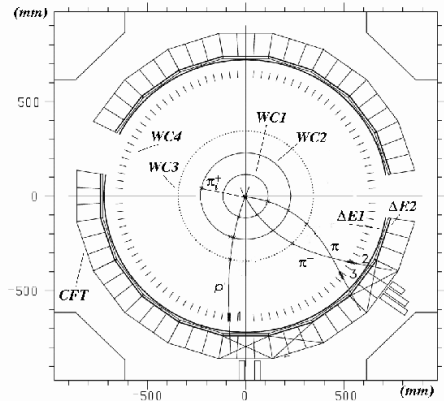}
  \setlength{\abovecaptionskip}{15pt}  
  \setlength{\belowcaptionskip}{5pt}  
  \caption{\footnotesize Reconstructed particle trajectories in CHAOS
    for $\pi^+_i \rightarrow \pi^+\pi^-p$ at T$_{\pi^+_i}$=300 MeV.
    The experimental layout shows the geometrical disposition of the 
    wire chambers (WC), the first level trigger hardware (CFT) and the 
    magnet return yokes in the corners. The CFT segments which are 
    hit by particles are marked with crosses, and the energy deposited
    in the first two layers ($\Delta E1$ and $\Delta E2$) is indicated 
    by boxes. $\Delta E1$ is hit by two pions, however the thinner 
    segmentation of $\Delta E2$ is sufficient to distinguish them.}
\end{figure}

telescope of fast counters (CFT). Fig. 1 shows the geometrical 
disposition of the magnet return yokes, wire chambers and 
telescope, and the trajectories of a reconstructed event, 
$\pi^{+}_{i}  \rightarrow \pi^+\pi^- p$ on $^{45}Sc$ at 
$T_{\pi^{+}}$=300 MeV. The magnetic field was varied with
$T_{\pi^{+}}$ to maintain constant the trajectory of the 
incident beam. The field was 0.55 T at $T_{\pi^{+}}$=240 MeV and 
0.69 at $T_{\pi^{+}}$=320 MeV. The inner and the outer WC's have 
a diameter of $\sim$23 cm and $\sim$123 cm, respectively. The 
latter is a vector WC, which operates in the fringe field of the 
magnet.  The overall WC efficiency to fully reconstructed
$\pi 2\pi$ events about 50$\pm$5\%. With this setup, the 
vector momentum $\vec{p_\pi}(p,\Theta)$ of a 130 MeV/c pion was 
analysed with $\delta$p$\sim 5$ MeV/c and 
$\delta \Theta \leq 2^{\circ} $. Both uncertainties are due mainly 
to pion multiple scattering. The CFT telescope encircles WC4. The 
CFT hardware is composed of two layers of NE110 plastic scintillator, 
0.3 cm and 1.2 cm thick respectively, and a layer of SF5 lead-glass 
of $\sim$5 radiation lengths. The telescope is segmented. Each 
segment covers an azimuthal angle $\Delta\Theta =18^\circ$. Segments 
can be removed to allow the pion beam to enter and exit CHAOS. The 
first CFT layer subtends the smallest zenith angle 
$\Delta\Phi$=$\pm 7^\circ$; therefore, it defines the geometrical solid 
angle of CHAOS $\Omega=1.5$ sr. The necessity of removing 2 (out of 20) 
CFT segments during a measurement leads to a nonuniform acceptance 
of CHAOS. In addition, pions decaying inside CHAOS further increase 
the irregularity of the acceptance. Thus, GEANT Monte Carlo 
simulations were required, which resulted in assigning a weight to 
each $\pi 2\pi$ event before it was binned. The weight distributions 
resemble those reported in Ref.\cite{expt:five} Fig. 4: they peak at 
{\em weight}$\sim$1.5 and monotonically decrease to zero at 
{\em weight}$\sim$7.  In order to avoid large corrections in the 
distributions, the soft cut $0< weight < \sigma +2\mu$ on reconstructed 
events is applied, where $\sigma$ and $\mu$ are the mean value and the 
standard deviation of the weight distributions, respectively. In the
case of T$_{\pi}$=300 MeV, $\sigma + 2\mu \sim$ 5.5.
 
2.3 Particle mass-identification 

Incident pions produce background reactions whose
rate overwhelms the $\pi 2\pi$ rate. Therefore, to study the 
$\pi 2\pi$ process in nuclei, a thorough identification of 
particle masses is required. In the case of CHAOS, this is
accomplished by binding the momentum (polarity) of a particle
to the pulse-height response of the telescope layers. In the 
present measurement, pion momenta do not exceed 230 MeV/c and 
the particles involved are $\pi$'s, $e$'s, $p$'s and $d$'s.

The pulse height response ($PH$) of pions and protons in 
plastic scintillators is $PH_{p}/PH_{\pi}> 5$ for p$_{\pi}\leq$ 
230 MeV/c see Fig. 8 in Ref. \cite{CHAOS:two} and Fig. 3 in 
Ref.\cite{expt:two}. By using only the first two CFT layers,
pions were separated from protons (deuterons) with high 
($\sim$100\%) selectivity. The same technique cannot favourably
be applied to $\pi$'s and $e$'s for momenta exceeding 110 MeV/c.  
Thus, SF5 lead-glass Cherenkov counters were used. 

Electrons (positrons) are byproducts of single charge exchange 
processes in the $^{45}Sc$ target: $\pi^+\rightarrow\pi^\circ$ 
followed by $\pi^\circ$ decay $\pi^\circ\rightarrow\gamma\gamma$ 
and $\gamma$ conversion $\gamma\rightarrow e^+e^-$. Thus, $e^+e^-$ 
pairs may look like $\pi^+\pi^-$ pairs. Fig. 2 shows the opening 
angle distributions of $e^+e^-$ ($\Theta_{e^+e^-}$, full squares) 
and $\pi^+\pi^-$ ($\Theta_{\pi^+\pi^-}$, full diamonds) pairs for 
the 300 MeV run,  where $\Theta$ is defined in the lab frame 
as the difference of the azimuthal angles of reconstructed pairs.
Most of the $\Theta_{e^+e^-}$ strength occurs near
0$^\circ$. An $e^+e^-$ to $\pi^+\pi^-$ misidentification 
would result in a contamination of $\Theta_{\pi^+\pi^-}$ at around 
0$^\circ$, and thus in a contamination of the $\pi^+\pi^-$ invariant 
mass at around the 2m$_\pi$ threshold (Ref. \cite{expt:five} Fig. 13). 
In order to safely separate electrons from pions, the following
strategy was followed. Soft kinematical cuts were assigned to reject 
those $\pi 2\pi$ spurious events whose momenta and total energy sum 
exceed the values allowed by the $^{45}Sc(\pi^+,\pi^+\pi^-p)^{44}Sc$ 
phase space. The same soft cuts were also applied to $e^+e^-$ pairs,
which brings about the distributions in Fig. 2.  The Cherenkov 
counters have a $\pi$ to $e$ discrimination efficiency above
95\% for momenta below 230 MeV/c (Ref. \cite{CHAOS:two} Fig. 11), 
which brings the $e^+e^-$ to $\pi^+\pi^-$ misidentification 
rate to about 0.25\%. This is further reduced to about 0.1\% by 
rejecting events with opening angles less than $3^\circ$.
For the distributions shown in Fig. 2, there are about 3000
$\Theta_{e^+e^-}$ events between 0$^\circ$ and 18$^\circ$, which 
is reduced to about 3 events by the CFT filtering combined with
the $3^\circ$ soft cut.  These events would mainly affect the 
$5^\circ$ $\Theta_{\pi^+\pi^-}$ datum which consists of about
900 events. Thus the surviving $e^+e^-$ contaminants have
\begin{figure}[htb]
 \centering
  \includegraphics*[angle=90,width=0.45\textwidth]{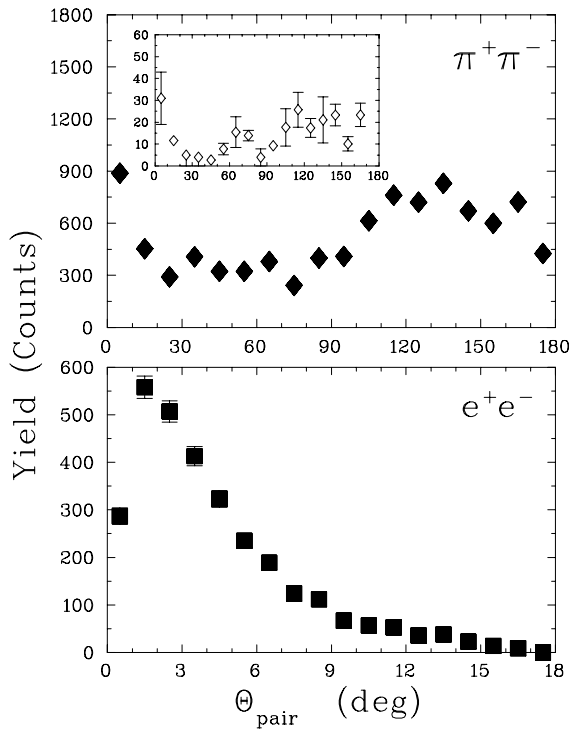}
  \setlength{\abovecaptionskip}{15pt}  
  \setlength{\belowcaptionskip}{5pt}  
  \caption{\footnotesize Opening angle distributions of $e^+e^-$ 
    (full squares) and $\pi^+\pi^-$ (full diamonds) pairs at 
    $T_{\pi^+}$=300 MeV. The inset diagram shows the $\pi^+\pi^-$ 
    opening angle distribution (open diamonds) for the 
    $\pi^+ \rightarrow \pi^+\pi^- p$ reaction; in this case, 
    $\pi^+\pi^-$ pairs are detected in coincidence with protons.}
\end{figure}
a negligible impact on the opening angle distributions, 
and in general on the $\pi 2\pi$ spectra. This feature is common 
to all the examined energies. In the upper panel, of Fig. 2, the 
$\Theta_{\pi^+\pi^-}$ distribution (open diamonds) is shown for pion 
pairs detected in coincidence with protons, namely, for pairs from 
the $\pi^+\rightarrow\pi^+\pi^- p$ reaction. In this case, kinematical 
cuts on the kinetic energy sum of particles detected in the reaction
exit channel (T$_{SUM}$) avoid contributions from $e^+e^-$ pairs in 
coincidence with a proton. Nevertheless, the two $\Theta_{\pi^+\pi^-}$ 
distributions show a similar behaviour. This second self-reliant result 
confirms that $e^+e^-$ events are efficiently rejected.

{\bf 3. Analysis} 

The $\pi 2\pi$ data are fully corrected within the geometrical
acceptance of CHAOS. No attempt to correct for data outside the 
CHAOS acceptance has been made, since this would involve relying 
either on phase space simulations or on model calculations. Such a 
{\em model-dependent} approach would alter the shape of distributions; 
in fact, the data must be extrapolated over a large $\Phi-$interval 
without the assistance of experimental data. This is illustrated in 
Fig. 3 for the $\pi\pi$ invariant mass distributions at T$_{\pi}$=280 
MeV: the full-line diagrams are the results of phase space simulations 
for the $^{45}Sc(\pi^+,\pi^+\pi^{\pm}N)^{44}X$ reactions for an 
ideal 4$\pi$ detector, while the shaded diagrams depict the effects
on distributions of the CHAOS geometry. In the present work, the shape 
of distributions will reflect the coplanar geometry of the spectrometer. 
\begin{figure}[htb]
 \centering
  \includegraphics*[angle=0,width=0.7\textwidth]{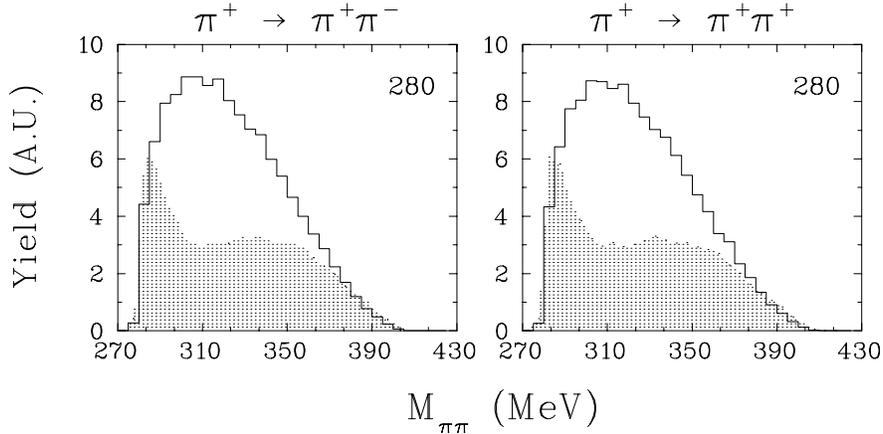}
  \setlength{\abovecaptionskip}{15pt}  
  \setlength{\belowcaptionskip}{5pt}  
  \caption{\footnotesize Results of phase space simulations for the 
    $^{45}Sc(\pi^+,\pi^+\pi^{\pm}N)^{44}X$ reactions at 280 MeV.
    Full-line diagrams: $\pi\pi$ invariant mass distributions 
    for an ideal 4$\pi$ detector; shaded diagrams: $\pi\pi$ 
    invariant mass distributions when including the CHAOS geometry.
    The diagrams are normalised at low and high invariant masses
    where distributions appear to have the same behaviour.}
\end{figure}

Once the particle mass is identified, the particle vector momenta are 
checked against simulations of the $^{45}Sc(\pi^+,\pi^+\pi^\pm N)^{44}X$ 
phase space, which include the CHAOS acceptance and resolution. 
A reconstructed event is rejected when it exceeds the allowed phase 
space volume. This procedure is applied to all the five energies. As 
an example, for the 300 MeV run the following soft cuts are applied: 
$p_{\pi^+}<$215 MeV/c, $p_{\pi^-}<$215 MeV/c and $T_{SUM}<$156 MeV.
As earlier mentioned, each $\pi 2\pi$ event is weighted before being 
binned. In order to avoid large corrections, the soft cut 
$0< weight < 5.5$ on reconstructed events is applied. In the case of 
the $\pi^{+} \rightarrow \pi^{+}\pi^{-}$ channel, the number of 
reconstructed events which passed the above tests are several 
thousands for each energy. The number drops by an order of magnitude 
for the $\pi^{+} \rightarrow \pi^{+}\pi^{+}$ channel.

\begin{figure}[htb]
 \centering
  \includegraphics*[angle=0,width=0.7\textwidth]{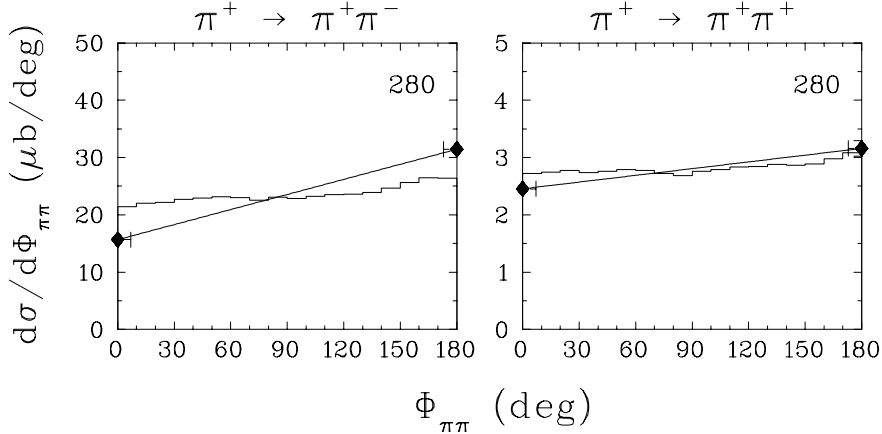}
  \setlength{\abovecaptionskip}{15pt}  
  \setlength{\belowcaptionskip}{5pt}  
  \caption{\footnotesize Out-of-plane dependence of  
    $d\sigma/d\Phi_{\pi\pi}$  as a function of $\Phi_{\pi\pi} \equiv 
    \Phi_{\pi_1\pi_2} = |\Phi_{\pi_1} - \Phi_{\pi_2 }|$ for the 
    $\pi^+\rightarrow\pi^+\pi^{\pm}$ reaction channels at 
    T$_{\pi^+}$=280 MeV. Full diamonds: experimental
    points; full lines: curves linearly joining the experimental data; 
    full histograms: results of phase-space simulations for the 
    $^{45}Sc(\pi^+,\pi^+\pi^{\pm}N)^{44}X$ reactions normalised 
    to the data at 0$^\circ \pm 7^\circ$ and 180$^\circ \pm 7^\circ$.}
\end{figure}
The data reduction is based on $\pi \rightarrow \pi_1\pi_2$ events 
which passed the following tests: have the origin in the target 
region, be mass- and charge-identified, survive the tests on 
kinematics and weights. Many-fold differential cross sections
$d^4\sigma/(dT d\Omega)_{\pi_1}(dT d\Omega)_{\pi_2}$ are then 
formed, where $T$ is the kinetic energy of a final pion and 
$\Omega$ is the solid angle into which a pion is scattered. 
$\Omega_\pi$ is related to the zenithal angle $\Phi_\pi$ and the 
azimuthal angle $\Theta_\pi$ via the equation 
$d\Omega_\pi$ = $d$cos$(\Theta_\pi) d\Phi_\pi$. The two angles 
reflect the coplanar geometry of CHAOS, which  restricts the 
out-of-plane acceptance to vary within $\pm 7^{\circ}$, while 
particles are accepted from $0^\circ$ to $360^\circ$ 
in the plane of the reaction except for two angular segments 
each $18^\circ$ wide. The latter account for the removal of two 
CFT's which allow the pion beam to enter and exit CHAOS (see Fig. 1). 
The data are represented as triple differential cross sections 
$d^{3}\sigma / d{\cal O}_{\pi\pi} d\Omega_\pi d\Omega_\pi$, where 
${\cal O}_{\pi\pi}$ represents $(T$ or $\Theta)_{\pi\pi}$ or a 
combination of them, and $\pi$ denotes a charged pion. 
$\frac{d^3\sigma}{d{\cal O}d\Omega d\Omega}$ is determined
by the quantities $f_e\frac{N({\cal O})}{\Delta{\cal O}}$, where $f_e$ 
is a parameter which is determined by the experimental conditions, 
$N({\cal O})$ is the number of weighted events in a given bin and 
$\Delta{\cal O}$ is the bin width for the ${\cal O}$ observable. 
The systematic uncertainty in assessing $f_e$ is 13.4\% ($\sigma$) 
at the central energy $T_{\pi^+}$=280 meV. A more detailed analysis 
for $T_{\pi^+}$=280 MeV is reported in Ref.\cite{expt:five}. The 
error bars reported on the $\pi 2\pi$ spectra account only for the 
statistical uncertainty. The overall uncertainty is obtained by 
summing the systematic and the statistical uncertainties in 
quadrature. In order to 
obtain the total cross section $\sigma_T$, the differential cross 
section $d^{3}\sigma / d{\cal O}_{\pi\pi} d\Omega_\pi d\Omega_\pi$ 
is summed over ${\cal O}_{\pi\pi}$ and $\Omega_\pi$'s. Fig. 4 
illustrates the out-of-plane dependence of the cross section 
$d\sigma/d\Phi_{\pi\pi}$ at 280 MeV, where $\Phi_{\pi\pi} 
\equiv \Phi_{\pi_1\pi_2} = |\Phi_{\pi_1} - \Phi_{\pi_2 }|$. The 
integration over the 
unmeasured portion of $\Phi_{\pi\pi}$ is accomplished either 
by using a linear function joining $\Phi_{\pi\pi}=0^{\circ}$ to 
$\Phi_{\pi\pi}=180^{\circ}$, or by interpolating the two measured 
points with the reaction phase space. Such an approach is applied
to all the incident pion energies and the results are similar. The 
values of $\sigma_T$ so obtained are listed in table 2 (Sec. 6.4). 

{\bf 4. The $\pi^+\rightarrow\pi^+\pi^-p$ exclusive measurement 
        at $T_{\pi^+}$=260 and 300 MeV.} 
 
At $T_{\pi^+}$=300 MeV kinetic energy, the $\pi 2\pi$ data were 
collected with sufficient statistics (over 10$^4$ reconstructed events) 
to enable an analysis of the $\pi^+(n)\rightarrow\pi^+\pi^-p$ exclusive 
reaction channel. Some results are discussed below.

Fig. 5 shows the missing energy ($E_M$) distribution of the 
$\pi^{+}$ $^{45}Sc \rightarrow \pi^{+} \pi^{-}p$ $^{44}Sc$ reaction 
at $T_{\pi^+}$= 260 and 300 MeV. The missing energy is defined
by the equation $E_M = [M_{^{45}Sc}-M_{^{44}Sc}-m_{\pi}-m_p]
           - [T^f_{\pi^+}+T^f_{\pi^-}+T^f_{p}-T^i_{\pi^+}]$, 
\begin{figure}[htb]
 \centering
  \includegraphics*[angle=0,width=0.25\textwidth]{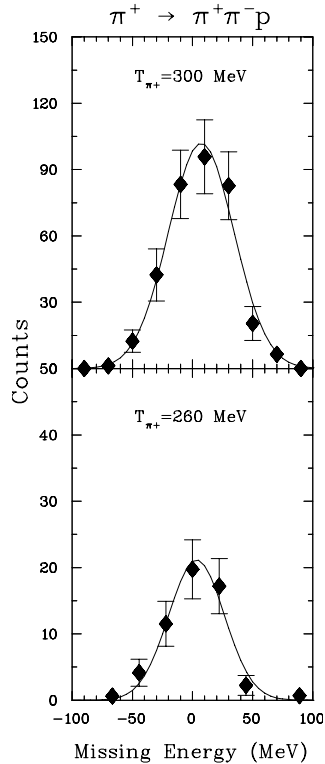}
  \setlength{\abovecaptionskip}{15pt}  
  \setlength{\belowcaptionskip}{5pt}  
  \caption{\footnotesize Missing energy distributions of the 
    $\pi^+$ $^{45}Sc \rightarrow \pi^+\pi^- p$ $^{44}Sc$ reaction at 
    incident kinetic energies of 260 and 300 MeV. The definition of
    missing energy is given in the text; in this case, $\pi^{+}\pi^{-}$ 
    pairs are detected in coincidence with protons. The curves are a 
    Gaussian fit to the data.}
\end{figure}
where the symbol $M (m)$ indicates a nucleus (particle) mass, while 
$i$ and $f$ denote the initial and final particles, respectively. 
 In this framework, $E_M$ represents the excitation energy of 
the $^{44}Sc$ residual system. The $E_M$ mean values, obtained with 
a Gaussian fit to the data, are $\sim$7 MeV at $T_{\pi^+}$=300 MeV, 
and $\sim$3 MeV at $T_{\pi^+}$=260 MeV. These values are consistent 
with $E_M\sim$0 MeV, when considering that the residual nucleus may 
be left in an excited state. Therefore, the $\pi \rightarrow \pi \pi$ 
reaction at intermediate energies is a quasifree process, which mainly 
involves a single nucleon $\pi N\rightarrow \pi \pi N$. A similar 
result was found in Ref.\cite{expt:two}, where the $\pi 2\pi$ reaction 
was studied as a function of the nuclear mass number.

Fig. 6 shows the invariant mass distributions (shaded diagrams) of 
$\pi^+\pi^{\pm}$ pairs that are detected without the requirement
of a proton in coincidence. 
The $M_{\pi\pi}$ distributions are also plotted as a 
function of $p_{\pi\pi} = |\vec{p}_{\pi^+} + \vec{p}_{\pi^\pm}|$, 
in order to illustrate the $p_{\pi\pi}-$dependence of $M_{\pi\pi}$ 
for the two 
\begin{figure}[htb]
 \centering
  \includegraphics*[angle=90,width=0.5\textwidth]{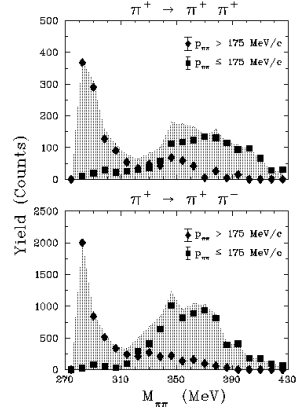}
  \setlength{\abovecaptionskip}{15pt}  
  \setlength{\belowcaptionskip}{5pt}  
  \caption{\footnotesize Invariant mass distributions of 
    $\pi^+\pi^{\pm}$ pairs at 300 MeV for $^{45}Sc$ as a function 
    of $p_{\pi\pi}$, the magnitude of the pion pair momentum. Shaded 
    diagrams $0 \leq p_{\pi\pi} \leq 350$ MeV/c (the maximum momentum 
    reached by a pion pair), full squares $0 \leq p_{\pi\pi} \le 175$ 
    MeV/c, and full diamonds $175 < p_{\pi\pi} \leq 350$ MeV/c.}
\end{figure}
reaction channels. The two channels have a similar behaviour: the 
low $M_{\pi\pi}$ bins are mainly populated by pairs (full diamonds) 
with the highest momenta $p_{\pi\pi}>$175 MeV/c. These pairs, 
however, show a different behaviour when their 
opening angle distributions are compared.
Fig. 7 depicts the $\Theta_{\pi\pi}$ 
\begin{figure}[htb]
 \centering
  \includegraphics*[angle=90,width=0.5\textwidth]{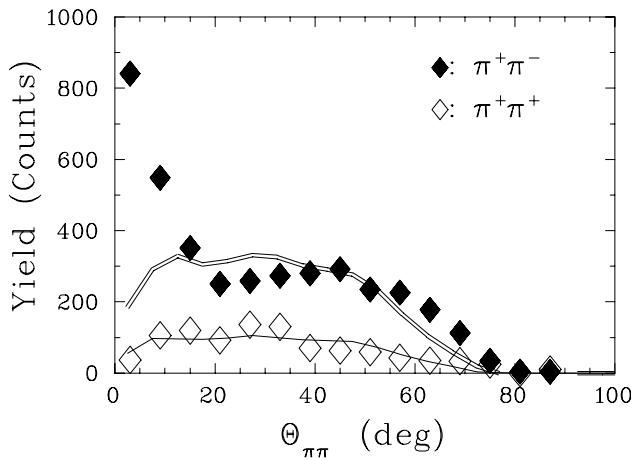}
  \setlength{\abovecaptionskip}{15pt}  
  \setlength{\belowcaptionskip}{5pt}  
  \caption{\footnotesize Opening angle distributions of $\pi^+\pi^+$ 
    and $\pi^+\pi^-$ pairs at 300 MeV, for $175 < p_{\pi\pi}\leq 350$ 
    MeV/c and  $2m_\pi \leq M_{\pi\pi} \leq 310$. The full and double
    lines are the results of phase space simulations for the 
    reactions $\pi^+$ $^{45}Sc \rightarrow\pi^+\pi^+ n$ $^{44}Ca$ 
    and $\pi^+$ $^{45}Sc \rightarrow\pi^+\pi^- p$ $^{44}Sc$, 
    respectively. Simulations include the CHAOS acceptance as well 
    as the $p_{\pi\pi}$ and $M_{\pi\pi}$ cuts, and are normalised 
    to the data from $\Theta_{\pi\pi} > 25^{\circ}$.}
\end{figure}
distributions for $p_{\pi\pi}>$175 MeV/c and 
$2m_\pi \leq M_{\pi\pi} \leq 310$ MeV. In this invariant mass 
range, $\pi^+\pi^+$ (open diamonds) and $\pi^+\pi^-$ (full 
diamonds) pairs are known to carry I=2 J=0 and I=0 J=0 quantum 
numbers, respectively \cite{expt:one,expt:five,ph_Dh:five,I=J=0:one}.
The $\Theta_{\pi^+\pi^+}$ distribution is well described by the
$\pi^+$ $^{45}Sc \rightarrow\pi^+\pi^+ n$ $^{44}Ca$ phase space 
(full line).  On the other hand, the $\Theta_{\pi^+\pi^-}$
distribution
displays a sharp peak at around $0^\circ$, before continuing as 
$\pi^+$ $^{45}Sc \rightarrow\pi^+\pi^- p$ $^{44}Sc$ phase space
(double line). For both channels, simulations include the CHAOS 
acceptance as well as the $p_{\pi\pi}$ and $M_{\pi\pi}$ cuts, and 
are normalised to the data from $\Theta_{\pi\pi} > 25^{\circ}$. 
In the case examined, the only difference between $\pi^+\pi^+$ 
and $\pi^+\pi^-$ pair is their quantum numbers, demonstrating
the strong influence of nuclear matter on 
the $(\pi\pi)_{I=J=0}$ interaction. Low $M_{\pi^+\pi^-}$ are 
populated by $(\pi\pi)_{I=J=0}$ pairs which decay preferentially
with small $\Theta_{\pi\pi}$ and high $p_{\pi\pi}$.

{\bf 5. Model of the $\pi 2\pi$ reaction in nuclear matter} 

The experimental results will be compared  with the model
predictions of Ref. \cite{ph_Dh:five}.
There are several good reasons for doing so.
The calculation of the $\pi 2\pi$ cross sections 
is based on a Monte Carlo method, which permits the CHAOS 
acceptance to be easily implemented in the code. The code 
requires the energy of the incident pion as an input parameter, 
thus the experimental distributions can be compared with the 
cross sections precisely at the same energies. For each 
accepted event, the code returns the vector momentum of each 
pion. A similar approach is followed in the data analysis,
facilitating the comparison between calculated cross 
sections and distributions of experimental observables.  

The model describes the elementary pion production reaction
by involving only one nucleon of the nucleus, via the quasifree 
process $\pi N\rightarrow \pi\pi N$ \cite{ph_Dh:five,ph_Dh:six}. 
The construction of the $\pi N\rightarrow \pi\pi N$ amplitude 
relies on one-point diagrams, which include both the pion-pole 
and contact term, and two- and three-point diagrams. Nucleons as 
well as the $\Delta(1232)$ and $N^\star(1440)$ can be excited as 
intermediate states. Standard nuclear effects, which may modify 
the reaction cross section, are accounted for.  These include Fermi
motion, Pauli blocking, pion absorption and quasielastic 
scattering. The role played by the nuclear medium on 
$(\pi\pi)_{I=J=0}$ interacting pairs is carefully examined. A
tangible modification of the $\pi\pi$ interaction is determined
by the coupling of pions to {\em p-h} and {\em $\Delta$-h} 
states, which enhances the intensity of the $\pi\pi$ spectral 
function at around the 2m$_\pi$ threshold. The effects of the
nuclear medium on the $\sigma$ meson properties are analysed
in separate articles \cite{sigma:six,sigma:seven}. The $\sigma$ 
mesons are dynamically generated by the rescattering of 
$(\pi\pi)_{I=J=0}$ pions. Medium modifications are studied 
by coupling $\sigma$'s to the nucleons by means of tadpole 
diagrams. Once inserted into the model, the tadpole terms are 
found to contribute negligibly. Therefore, the only detectable 
nuclear medium modification on the $(\pi\pi)_{I=J=0}$ system 
comes from the P-wave coupling of pions to {\em p-h} and 
{\em $\Delta$-h} configurations. 
\begin{figure}[htb]
 \centering
  \includegraphics*[angle=90,width=0.6\textwidth]{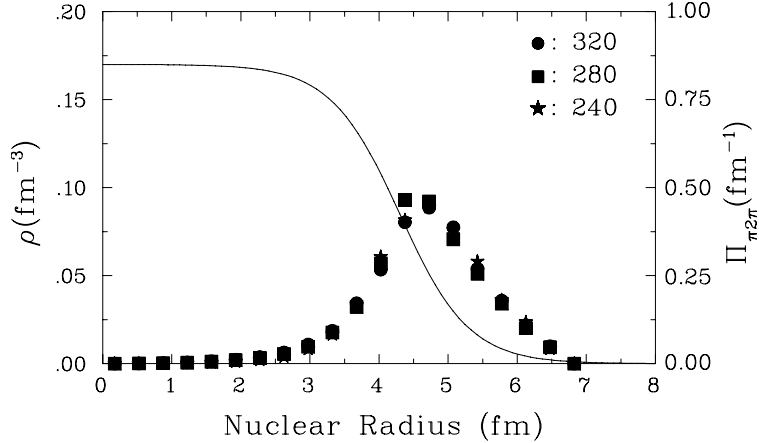}
  \setlength{\abovecaptionskip}{15pt}  
  \setlength{\belowcaptionskip}{5pt}  
  \caption{\footnotesize Nuclear density distribution ($\rho$, full 
    curve) as a function on the nuclear radius for $^{45}Sc$. The 
    probability of a $\pi 2\pi$ event to occur ($\Pi_{\pi 2\pi}$) is 
    described by the distributions of the solid points: circles 
    320 MeV (energy of incident pions), squares 280 MeV and stars 
    240 MeV. The $\Pi_{\pi 2\pi}$ distributions are normalized to unity, 
    the $\Pi_{\pi 2\pi}$ and $\rho$ yields are calculated with the 
    model of Ref. \cite{ph_Dh:five}.}
\end{figure}
The model is able to predict the probability distribution of an 
interesting observable $\Pi_{\pi 2\pi}$, which cannot be measured. 
$\Pi_{\pi 2\pi}$ is the composite probability of a $\pi 2\pi$ event 
to take place and be detected (by CHAOS) as a function of the 
nuclear radius. Fig. 8 shows the distributions for various energies 
of incident pions: circles 320 MeV, squares 280 MeV and stars 240 
MeV. The $\Pi_{\pi 2\pi}$ distributions, when normalized to unity, 
vary slightly with energy thus indicating minor variations of the 
(average) nuclear density at which a $\pi 2\pi$ event occurs. In 
fact, $\rho$=0.36$\rho_n$ at T$_{\pi^+}$=320 MeV, $\rho$=0.37$\rho_n$ 
at T$_{\pi^+}$=280 MeV, and $\rho$=0.35$\rho_n$ at T$_{\pi^+}$=240 MeV.

{\bf 6. The T-dependence of the 
      $\pi^+ \rightarrow \pi^+\pi^{\pm}$ reactions in $^{45}Sc$} 

The experimental results are compared with the predictions of the
model described in Sec. 5, and with the phase space of the 
$\pi^+$ $^{45}Sc \rightarrow \pi^+\pi^{\pm} N$ $^{44}X$ reactions. 
Both model and phase space take the $\pi 2\pi$ CHAOS acceptance
into account. 
The error bars reflect statistical uncertainties. Systematic 
uncertainties for T$_{\pi^+}$=280 MeV were discussed in Sec. 3.

6.1 The $\pi\pi$ invariant mass, M$_{\pi\pi}$

In order to check the consistency of the results, the 
M$_{\pi\pi}^{Sc}$ distribution at 280 MeV is compared to 
the published M$_{\pi\pi}^{Ca}$ cross section at 283 MeV
\cite{expt:five}. The comparison is shown in Fig. 9 top panels 
for the $\pi^+ \rightarrow \pi^+\pi^{\pm}$ reaction channels. 
The invariant mass of Sc (full diamonds) is normalised to the Ca 
invariant mass (open diamonds): as expected, the two distributions 
display similar shapes since both kinetic energy and nucleus mass 
number are close. 
\begin{figure}[htb]
 \centering
  \includegraphics*[angle=0,width=0.6\textwidth]{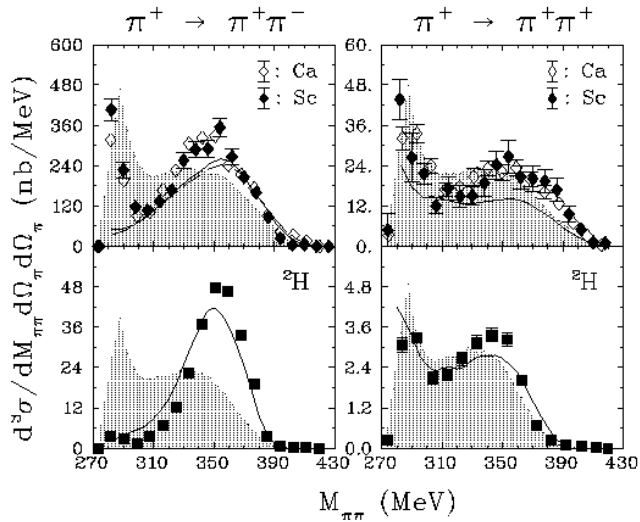}
  \setlength{\abovecaptionskip}{15pt}  
  \setlength{\belowcaptionskip}{5pt}  
  \caption{\footnotesize Invariant mass distributions of pion pairs 
    for the $\pi^+ \rightarrow \pi^+\pi^{\pm}$ reaction channels. 
    The Ca data (open diamonds, \cite{expt:five}) were taken at 283 
    MeV and are used for normalisation. The Sc M$_{\pi\pi}$ data
    are indicated with full diamonds and were taken at 280 MeV. The 
    phase space of the $^{45}Sc(\pi^+,\pi^+\pi^{\pm}N)^{44}X$ 
    reactions are represented by the shaded diagrams. The curves 
    are from \cite{ph_Dh:five,sigma:six,sigma:seven}. The continuous 
    lines denote the results of the full model. The dashed line in the 
    upper left panel is obtained when pions are coupled only to {\em p-h} 
    and {\em $\Delta$-h} states. More details are given in the text.}
\end{figure}
\begin{figure}[htb]
 \centering
  \includegraphics*[angle=0,width=0.50\textwidth]{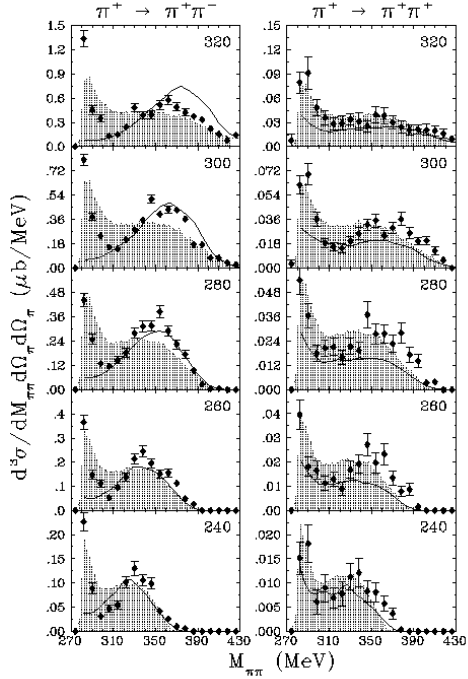}
  \setlength{\abovecaptionskip}{15pt}  
  \setlength{\belowcaptionskip}{5pt}   
  \caption{\footnotesize Invariant mass distributions of pion pairs
    as a function of the incident pion kinetic energy. The phase space
    for $^{45}Sc(\pi^+,\pi^+\pi^{\pm}N)^{44}X$ reactions is indicated 
    by the shaded diagrams. The full lines are the model calculations
    of \cite{ph_Dh:five,sigma:six,sigma:seven}. The kinetic energy 
    of the pion beam is given in each panel.}
\end{figure} 
The data (full squares) in the lower panels are the $\pi\pi$ 
invariant mass distributions in deuterium, namely, of the 
elementary reactions $\pi^+ n(p) \rightarrow \pi^+ \pi^- p(p)$ 
and $\pi^+ p(n) \rightarrow \pi^+ \pi^+ n(n)$ \cite{expt:five}. 
The data are compared with the model predictions (full line) 
described in Sec. 5 Ref. \cite{ph_Dh:five}. The noticeable 
feature is that, for the reactions on $^{2}$H,
the calculations are able to reproduce the 
varying behaviour of the M$_{\pi\pi}$ distributions at around 
threshold: M$_{\pi^+\pi^-}$ is weakened to nearly zero while 
M$_{\pi^+\pi^+}$ reaches its maximum. 

In nuclei (Ca or Sc,  upper panels) the situation changes. The
M$_{\pi^+\pi^+}$ distributions resemble that of the elementary 
reaction and are well explained by the model predictions (solid
line). In the $\pi^+ \rightarrow \pi^+\pi^-$ channel, M$_{\pi\pi}$ 
displays a remarkable reshaping with respect to the elementary
process, and the fully renormalised calculations (solid line) give 
only a partial account of the threshold enhancement of the data. 
Furthermore, the invariant mass intensity decreases slightly at 
around threshold when only the P-wave coupling of pions to 
{\em p-h} and {\em $\Delta$-h} states is considered (dashed line). 

A further element of comparison is the reaction phase space (shaded 
diagrams in Fig. 9), which is normalised to data. Phase space
describes the M$_{\pi^+\pi^+}$ distributions throughout the energy 
range, but largely misses the M$_{\pi^+\pi^-}$ distributions. 

The features observed for Sc at 280 MeV are in general observed at
all the examined energies. This is illustrated in Fig. 10,
which shows the M$_{\pi^+\pi^\pm}$ cross sections as a function of 
the incident pion
kinetic energy. The $\pi^+\pi^+$ invariant mass cross section
is well explained by a microscopic model of the pion production 
reaction \cite{ph_Dh:five}, and the distributions are also described 
by the $\pi 2\pi$ phase space.  This leads to the following 
conclusions. The dynamical traits of the $(\pi\pi)_{I=2J=0}$ 
interaction are understood, whether it occurs on the nucleon or in 
nuclei. Furthermore, the nuclear medium does not appreciably change 
the $\pi\pi$ interaction in the I=2 J=0 channel with respect to the 
vacuum. 

\begin{figure}[h]
 \centering
  \includegraphics*[angle=0,width=0.50\textwidth]{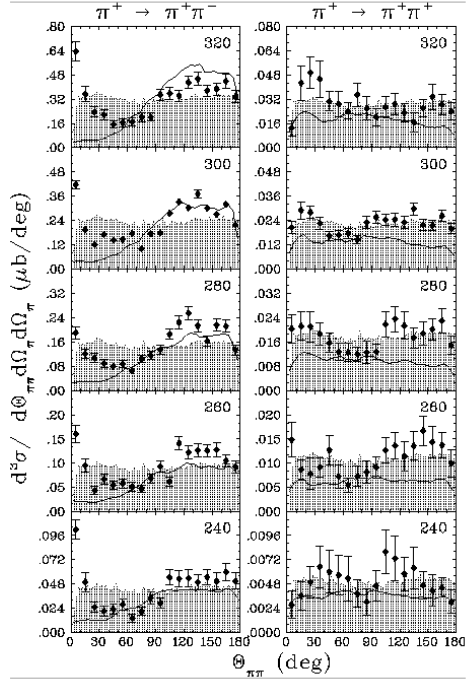}
  \setlength{\abovecaptionskip}{15pt}  
  \setlength{\belowcaptionskip}{5pt}   
  \caption{\footnotesize Opening angle distributions of pion pairs
    as function of the incident pion
    kinetic energy (full diamonds). The phase space 
    of the $^{45}Sc(\pi^+,\pi^+\pi^{\pm}N)^{44}X$ reactions is indicated 
    by the shaded diagrams. The full lines are the model calculations
    of \cite{ph_Dh:five,sigma:six,sigma:seven}. The pion incident 
    energy is given in each panel.}
\end{figure}

In the $\pi^+ \rightarrow \pi^+\pi^-$ channel, the terms describing 
the $\pi N \rightarrow \pi\pi N$ amplitude must largely cancel out 
to explain the threshold depletion present in the model calculations.
This view is also corroborated
by the phase space behaviour, which approaches its maximum at around 
threshold. The enhancement measured at different kinetic energies 
(and nuclei) is only partly explained by theory, thus implying that 
either the cancellations are no longer effective in nuclear matter,
or the $\sigma$ (i.e., strongly interacting $(\pi\pi)_{I=J=0}$ pairs) 
and its coupling to nucleons is not properly accounted for. 
In fact, other models of the $(\pi\pi)_{I=J=0}$ interaction in
nuclear matter \cite{sigma:one,sigma:two} find that the appearance 
of the $\sigma$ largely contributes to heighten the $\pi\pi$ spectral 
function at threshold. These models, however, do not embed any pion 
production process. As a final comment, it is worthwhile noting that
all the $\pi 2\pi$ data are well reproduced by both theory and phase 
space over the high invariant mass range.

6.2 The $\pi\pi$ opening angle, $\Theta_{\pi\pi}$

For a reconstructed pion pair $\pi_1 \pi_2$, the opening 
angle is defined as 
$\Theta_{\pi_1\pi_2}$ = $\Theta_{\pi_1}$ - $\Theta_{\pi_2}$, where 
the azimuthal angles $\Theta_{\pi_1}$ and $\Theta_{\pi_2}$ are 
measured in the lab system. The indices 1 and 2 are assigned such 
that $\Theta_{\pi_1\pi_2}$ is restricted to vary from 0$^\circ$ to 
180$^\circ$. The $\pi\pi$ opening angle is an observable that proved 
to be sensitive to the isospin state of pion pairs (see Sec. 4 Fig. 7).
The $\Theta_{\pi\pi}$ distributions are presented in Fig. 11 for 
all the examined energies, and compared with the reaction 
theory and phase space. In the $\pi^+\rightarrow\pi^+\pi^+$ channel, 
the $\Theta_{\pi\pi}$ distributions are generally flat over the 
angular interval 0$^{\circ}$-180$^{\circ}$. Such behaviour is 
predicted by the theory and phase space. The $\Theta_{\pi^+\pi^-}$ 
distributions are peaked at around $\sim 0^{\circ}$, regardless of 
the energy. Such a feature is not described by the model calculations, 
which predict a smooth and decreasing intensity of $\Theta_{\pi\pi}$ 
when approaching zero degrees. For the sake of comparison, in Fig. 12,
the $\Theta_{\pi\pi}$ distributions for the elementary 
reactions $\pi^+ n(p) \rightarrow \pi^+ \pi^- p(p)$ and 
$\pi^+ p(n) \rightarrow \pi^+ \pi^+ n(n)$ at T$_{\pi^+}$=283 MeV
\cite{expt:three,expt:five} are shown. $\Theta_{\pi^+\pi^-}$ is 
characterised by a negligible strength below $\sim 60^{\circ}$. 
\begin{figure}[htb]
 \centering
  \includegraphics*[angle=0,width=0.6\textwidth]{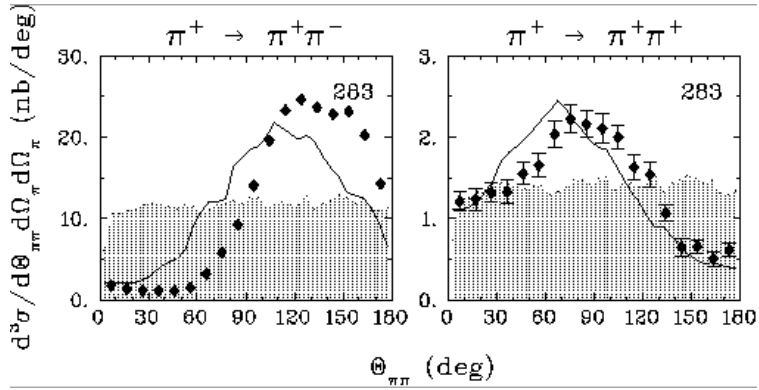}
  \setlength{\abovecaptionskip}{15pt}  
  \setlength{\belowcaptionskip}{5pt}   
  \caption{\footnotesize Opening angle distributions at 283 MeV 
    kinetic energy. The shaded diagrams denote the phase space of 
    the $^{2}H(\pi^+,\pi^+\pi^{\pm}N)N$ reactions. The full lines 
    are the model calculations of Refs. 
    \cite{ph_Dh:five,sigma:six,sigma:seven}.}
\end{figure}
In this angular span, the strength detected in nuclei is a direct 
consequence of the medium modifications on the $(\pi\pi)_{I=J=0}$ 
interaction, since the $\pi N\rightarrow \pi \pi N$ reaction is a 
quasifree process. The $\pi^+ \rightarrow \pi^+ \pi^+$ channel
behaves differently. The elementary reaction already exhibits 
strength below 60$^\circ$, which does not increase in Sc. Theory 
\cite{ph_Dh:five} is able to predict both the intensity and the 
shape of the $\Theta_{\pi^+\pi^+}$ distributions for Sc.  For 
$^2$H, calculations account correctly for the measured intensities
over the interval  $0^\circ \le \Theta_{\pi^+\pi^+} < 180^\circ$, 
but the distribution is predicted to peak at about 60$^\circ$, 
20$^\circ$-25$^\circ$ below the measured value.

For the $\pi\pi$ opening angle analysis, the conclusions that can be 
drawn are similar to those for the invariant mass. At intermediate 
energies, nuclear matter clearly modifies the $(\pi\pi)_{I=J=0}$ 
interaction, while it leaves the $(\pi\pi)_{I=2J=0}$ interaction rather 
unaltered. The theory is able to explain the $\pi 2\pi$ process only 
when it occurs in the I=2 J=0 channel.

6.3 The $\pi$ kinetic energy, T$_\pi$.

The observable under consideration, and illustrated in Fig. 13,
is the kinetic energy of a single 
pion. In the $\pi^+ \rightarrow \pi^+\pi^-$ channel, positive (full 
\begin{figure}[htb]
 \centering
  \includegraphics*[angle=0,width=0.50\textwidth]{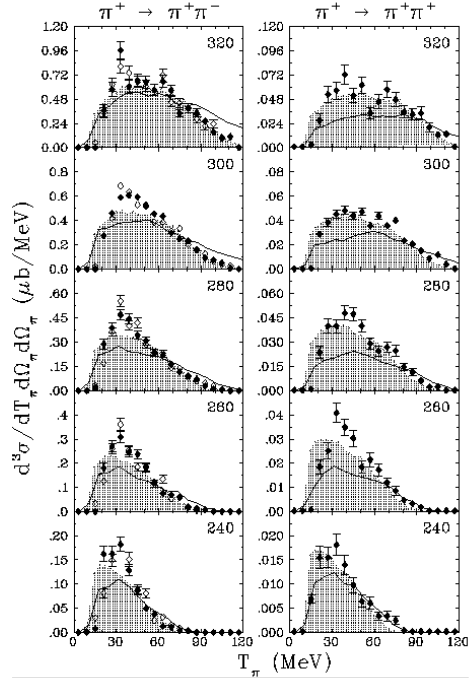}
  \setlength{\abovecaptionskip}{15pt}  
  \setlength{\belowcaptionskip}{5pt}   
  \caption{\footnotesize Single pion kinetic energy distributions as
    a function of the incident pion kinetic energy. Positive pion
    distributions are indicated with full diamonds, negative pions with 
    open diamonds. The shaded diagrams are the result of phase space 
    simulations for the $\pi^+$ $^{45}Sc \rightarrow \pi^+\pi^{\pm} N$ 
    $^{44}X$ reactions. The full lines are the positive pion energy
    distributions which are derived from the model of 
    Refs.  \cite{ph_Dh:five,sigma:six,sigma:seven}.}
\end{figure}
diamonds) and negative (open diamonds) pion distributions can barely 
be distinguished throughout the energy range examined. The full lines
\cite{ph_Dh:five} describe the positive pion energy distributions. 
Calculations provide a fair description of the measured distributions,
but tend to overestimate the intensities of the data in the
high-energy range. They predict greater abundances of energetic pions, 
which are also in excess of those expected from
$\pi^+$ $^{45}Sc \rightarrow \pi^+\pi^- p$ $^{44}Sc$ phase space 
(shaded diagram). The $T_{\pi^+}$ distributions are well accounted 
for by the $\pi^+$ $^{45}Sc \rightarrow \pi^+\pi^+ n$ $^{44}Ca$ phase 
space simulations, which are normalise to the $\pi 2\pi$ data.  On 
the theoretical side, calculations for the
$\pi^+ \rightarrow \pi^+\pi^+$ channel
provide a degree of agreement with 
data similar to that found for the $\pi^+ \rightarrow \pi^+\pi^-$ 
channel.
 
The study of the T$_\pi$ observable completes the present discussion.
It confirms the framework already sketched out by the study of the
previous observables, i.e. M$_{\pi\pi}$ and $\Theta_{\pi\pi}$. The
$\pi^+ \rightarrow \pi^+\pi^+$ reaction channel can be explained
in terms of both model calculation and phase space. The 
$\pi^+ \rightarrow \pi^+\pi^-$ data are only partially clarified;
in the present case, the model predictions fail to describe the 
high-energy tail of the T$_\pi$ distributions, which follow phase space. 

6.4 The $\pi 2\pi$ total cross sections, $\sigma_T$

With the CHAOS spectrometer one can directly determine 
$d^{3}\sigma / d{\cal O}_{\pi\pi} d\Omega_\pi d\Omega_\pi$, where 
${\cal O}_{\pi\pi}$ represents a generic observable, $d\Omega_\pi$
is the solid angle into which a  charged pion is scattered and 
$d\Omega_\pi$ = $d$cos$(\Theta_\pi) d\Phi_\pi$. In order to assess
$\sigma_T$, the multiple differential cross section has to be summed 
over ${\cal O}_{\pi\pi}$ and $\Omega_\pi$'s. The limited 
$\Phi$-acceptance of CHAOS, however, allows the determination of 
$d\sigma/d\Phi_{\pi\pi}$ only at  $\Phi_{\pi\pi}=0^{\circ}$ and 
$\Phi_{\pi\pi}=180^{\circ}$, where $\Phi_{\pi\pi} \equiv 
\Phi_{\pi_1\pi_2} = |\Phi_{\pi_1} - \Phi_{\pi_2 }|$.. Within this 
interval, the $\Phi-$dependence of the $\pi 2\pi$ reaction is 
presently assumed either to behave linearly, or to follow  the
$^{45}Sc(\pi^+,\pi^+\pi^{\pm}N)^{44}X$  phase space (see Sec. 3, 
and Fig. 4 for details). In this framework, the results obtained 
for $\sigma_T$ are those reported in Table 2. In 
the last two columns, the ratios $\sigma_T^{+-}/\sigma_T^{++}$ for 
the two reaction channels are listed. For a given T$_{\pi^+}$, the 
ratios for Sc and H agree within the error bars, which implies that 
a similar reaction mechanism underlies the $\pi 2\pi$ reaction 
whether it occurs in the nucleon or in nuclei. This confirms that 
the  $\pi 2\pi$ reaction in nuclei is a quasifree process.

\begin{table}[tcb]
\caption[Table]{
  Experimental total cross sections  of the 
  $\pi^{+} \rightarrow \pi^{+}\pi^{-}$ ($+-$) and
  $\pi^{+} \rightarrow \pi^{+}\pi^{+}$ ($++$) reaction channels on 
  $^{45}$Sc as a function of the pion incident energy T$_{\pi^+}$. 
  The uncertainties associated to the $phase-space$  cross 
  sections are omitted since they do not exceed the uncertainties
  of the $linear$ cross sections. In the last two columns the 
  $\sigma_T^{+-} / \sigma_T^{++}$ ratios are reported for Sc and H, 
  respectively.}
\begin{center}
  \begin{tabular}{cccccccc}                                          \hline 
    \multicolumn{1}{c}{T$_{\pi^+} [MeV]$   }            & 
    \multicolumn{4}{c}{$\sigma_T$ [$\mu b]$}            &
    \multicolumn{1}{c}{ }                               &
    \multicolumn{2}{c}{$\sigma_T^{+-} / \sigma_T^{++}$}           \\ 
    \multicolumn{1}{c} { }                          & 
    \multicolumn{2}{c} {$linear$}                   &   
    \multicolumn{2}{c} {$phase$ $space$}            &
    \multicolumn{1}{c} { }                          &  
    \multicolumn{2}{c} { }                                        \\ 
    \cline{2-5} \vspace{-1pc}    &     &     &     &     &    &   \\ 
 & $+-$ & $++$ & $+-$ & $++$ & & $^{45}$Sc & $^{1}$H              \\ \hline
  240 & 1225$\pm$164 & 143$\pm$19 & 1209 & 141 &  & 
             8.6$\pm$1.6& 7.1$\pm$1.2\cite{expt:nine}            \\ 
  260 & 2595$\pm$348 & 347$\pm$47 & 2559 & 341 &  &
             7.5$\pm$1.4& 6.1$\pm$0.9\cite{expt:nine}            \\ 
  280 & 4263$\pm$571 & 507$\pm$68 & 4233 & 505 &  &
             8.4$\pm$1.6& 7.4$\pm$1.0\cite{expt:nine}            \\ 
  300 & 6660$\pm$892 & 640$\pm$86 & 6643 & 637 &  &
            10.4$\pm$2.0& 9.6$\pm$1.5\cite{expt:nine}            \\ 
  320 & 9160$\pm$1227& 826$\pm$110& 9148 & 824 &  &
            11.1$\pm$2.1&12.4$\pm$3.0\cite{expt:ten}             \\ \hline
\end{tabular}
\end{center}
\end{table}

{\bf 7. Existing $\pi 2\pi$ results} 

Two collaborations have recently presented their studies of 
$\pi\pi$ dynamics in nuclear matter in proximity of the 2m$_\pi$ 
threshold. These are the Crystal Ball (CB) Collaboration at the 
AGS and the TAPS collaboration at MAMI. Comparisons can only be
made with the general features of the data, since pion pairs are 
produced by different reactions, or different isospin channels are 
studied, or different energies and nuclei are examined in the
different experiments.

The CB Collaboration has presented results on the 
$\pi^- A \rightarrow \pi^{\circ}\pi^{\circ} A'$ reaction for A: 
$H$, $^{12}C$, $^{27}Al$ and $^{64}Cu$, at an incident pion energy 
of $T_{\pi^-}$=291.6 \cite{expt:six}. Neutral pion pairs have I=0, 
therefore they can be directly compared to $\pi^+\pi^-$ pairs
\cite{expt:one,expt:five,ph_Dh:five,I=J=0:one}. A recent article 
\cite{expt:seven} compares the CB and CHAOS data, pointing out the 
relevant common features shared by the two data sets. The observable 
$\cal C$$_{\pi\pi}^A$ is of particular interest in the comparison. 
$\cal C$$_{\pi\pi}^A$ is proportional to the ratio of M$_{\pi\pi}^A$ 
to M$_{\pi\pi}^H$ \cite{expt:four,expt:five}. Consideration of such 
a ratio lessens the importance of the difference in the CB ($\sim$93\%  
of 4$\pi$) and CHAOS ($\sim$12\% of 4$\pi$) acceptances. Comparison 
shows that the behaviour of $\cal C$$_{\pi^{\circ}\pi{^\circ}}^C$ and 
$\cal C$$_{\pi^+\pi^-}^C$ is similar over the entire energy range.

$\pi\pi$ dynamics in nuclear matter has been also studied by 
the TAPS Collaboration. TAPS has used the photoproduction reaction 
$\gamma \rightarrow \pi\pi$ on H, $^{12}$C and $^{208}$Pb, at photon 
energies from 400 to 460 MeV\cite{expt:eight}. At these energies, 
gammas can penetrate to the interior of nuclei, therefore probing 
nuclear densities ($\rho\sim 0.35\rho_n$ for $^{12}$C and 
$\rho\sim 0.65\rho_n$ for $^{208}$Pb) higher than the pions in the 
present work ($\rho\sim 0.36\rho_n$). In the case of TAPS, medium 
effects on the I=0 $\pi\pi$ interacting system are understood to be 
broader than the CHAOS ones, but the nuclear distortions on the two 
pions being detected are also broader, which has an impact on the 
M$_{\pi\pi}^A$ cross section. Therefore, a simple comparison of 
$\cal C$$_{\pi^{\circ}\pi{^\circ}}^A$ to $\cal C$$_{\pi^+\pi^-}^A$ 
may be misleading, since distortions do not affect M$_{\pi\pi}^H$. 
TAPS has redefined the composite ratio 
$\cal C$$_{\pi\pi}$ $\equiv$ M$_{\pi\pi}^{Pb}$/M$_{\pi\pi}^C$. With 
this new definition, the TAPS $\cal C$$_{\pi\pi}$ agrees well with 
the CHAOS $\cal C$$_{\pi\pi}$ (Ref. \cite{expt:eight}). In addition,
strong medium modifications have been observed only for the 
$\gamma \rightarrow \pi^{\circ}\pi^{\circ}$ reaction channel; that is, 
for pion pairs interacting in the I=J=0 channel. Medium effects on 
the isospin I=1 $\pi\pi$ pairs have not been detected. Further points 
of contact between the TAPS and the CHAOS results are discussed in 
Ref. \cite{expt:eight}. 

{\bf 8. Conclusions} 

This article presents the results of an exclusive measurement of 
the pion-production $\pi^+ \rightarrow \pi^+\pi^{\pm}$ reactions on 
$^{45}Sc$, at incident pion energies of 240, 260, 280, 300 and 320 
MeV. In previous measurements the same reaction has been examined
on $^{2}H$, $^{4}He$, $^{12}C$, $^{16}O$, $^{40}Ca$ and $^{208}Pb$ 
nuclei, at an incident pion energy of $\sim$280 MeV \cite{expt:one,
expt:1.5,expt:two,expt:2.5,expt:three,expt:four,expt:five}. 
A two-fold interest has been driving these measurements:
understanding the general traits of the pion-production reaction in 
nuclei, and studying the $\pi\pi$ dynamics in nuclear matter. To 
this end, the CHAOS magnetic spectrometer at TRIUMF has been used, 
which permitted the analysis of $\pi^+\pi^+$ and $\pi^+\pi^-$ pairs 
simultaneously. The study of the medium modifications of the
$\pi^+\pi^-$ interacting system is of central importance, being 
the lightest system which carries the I=0 J=0 quantum numbers, that 
is, the quantum numbers of the QCD vacuum. At the present energies,
the I=2 pion pairs interact very weakly, thus, only weak medium
modifications on the $\pi^+\pi^+$ system are expected. Therefore,
the comparison of the two isospin channels yields direct indications 
of the nuclear medium effects on the $(\pi\pi)_{I=J=0}$ system, 
that is, on the $\sigma$ meson. 

The A- and T-dependence studies of the $\pi 2\pi$ reaction lead to 
the following conclusions. 
\begin{enumerate}
\item The $\pi 2\pi$ reaction in nuclei is a quasifree process, 
  which involves a single nucleon $\pi N \rightarrow \pi \pi N$.
\item An analysis of the p$_{\pi\pi}-$dependence of M$_{\pi\pi}$ 
  reveals that near the 2m$_\pi$ threshold $\pi\pi$ pairs carry 
  the highest momenta available, and the 
  opening angle distribution of $\pi^+\pi^-$ pairs is peaked at 
  $\sim$0$^\circ$. This behavior is not explained by the 
  reaction phase space or by theory.
\item The $\pi 2\pi$ data are compared with a complete model of 
  the pion-production reaction\cite{ph_Dh:five}.  At the present 
  energies, calculations indicate that the average nuclear density 
  at which the process occurs is $\rho \sim 0.36 \rho_n$, that is, 
  the process is confined to the nuclear skin.
\item The observables M$_{\pi\pi}$, $\Theta_{\pi\pi}$ and T$_\pi$ 
  have been carefully examined and compared to model calculations 
  and reaction phase space, both of which take the CHAOS acceptance
  into account. 
  The dynamics of the $\pi^+ \rightarrow \pi^+ \pi^+$ channel is 
  well explained by theory. Phase space provides a good description
  of the kinematic behaviour of the reaction over a 
  wide range of energies.  The nuclear medium does not 
  detectably modify the $(\pi\pi)_{I=2J=0}$ interaction. For 
  example, the threshold behavior of M$_{\pi^+\pi^+}$ depends 
  negligibly on A and T.  This also indicates that pion final 
  state interactions cannot account for the enhancement observed 
  at around the 2m$_\pi$ threshold. The dynamics of the 
  $\pi^+ \rightarrow \pi^+ \pi^-$ channel is far more varied. 
  For the elementary $\pi N \rightarrow \pi\pi N$ reaction, the 
  M$_{\pi^+\pi^-}$ strength at around threshold is depleted 
  over a wide range of incoming pion energies, from 223
  to 305 MeV Ref. \cite{expt:nine} and this work Fig. 9. Thus the 
  terms forming the elementary amplitude must cancel out. 
  In nuclear matter, M$_{\pi^+\pi^-}$ rapidly gains strength, 
  which is not caused by the final state interactions of pions 
  with the residual nucleus. Therefore, either the medium removes 
  cancellations or it produces new events. According to the authors 
  of Refs. \cite{sigma:one,sigma:two}, an event is the appearance 
  of the $\sigma$ meson, a genuine QCD system formed by a $q\bar{q}$
  state, which largely reshapes its spectral function in response to 
  nuclear matter.  The M$_{\pi^+\pi^-}$ threshold strength is 
  primarily due to this occurrence. However, this basic 
  result is not corroborated by a model of the $\pi 2\pi$ reaction. 
  The model used in this article accounts for the pion-production 
  process \cite{ph_Dh:five}. At around threshold, M$_{\pi^+\pi^-}$ 
  builds up strength via the P-wave coupling of pions to {\em p-h} 
  and {\em $\Delta$-h} nuclear excitations. This process also 
  underlays the dynamical origin of the $\sigma$ meson: a $\pi\pi$ 
  interacting state in the I=J=0 channel. This resonant state is 
  found to drop in mass and width at finite densities, i.e. at
  $\rho<\rho_n$ \cite{ph_Dh:seven}. However, the $\sigma$ strength 
  at around threshold is insufficient to explain the data. 
  This specific issue deserves the attention of theorists, since it 
  is of primary importance in the understanding of whether the invariant 
  mass shift toward threshold is due to the partial restoration of 
  chiral symmetry or to other essential events. Standard nuclear 
  effects are already accounted for by the model used in the 
  present analysis. This general conclusion also applies to the  
  $\Theta_{\pi\pi}$ and T$_\pi$ observables. 
\item The $\pi 2\pi$ results obtained by the CHAOS Collaboration 
  have been compared to the results of the CB and the TAPS 
  Collaborations, which are the only available ones. The results 
  discussed in the CHAOS articles share relevant common features 
  with the results presented by these two collaborations. The
  discussion over the composite ratio $\cal C$$_{\pi\pi}$ $\equiv$ 
  M$_{\pi\pi}^A$/M$_{\pi\pi}^H$ has been only briefly developed 
  since it is the subject of a forthcoming article.
\end{enumerate}

{\bf Acknowledgments} 

The authors would like to acknowledge the support received from TRIUMF. 
The present work was made possible by grants from the Istituto Nazionale 
di Fisica Nucleare (INFN) of Italy, the National Science and Engineering 
Research Council (NSERC) of Canada, the Australian Research Council and
the German Ministry of Education and Research (BMBF 06TU987). The 
authors would also like to acknowledge useful discussions with 
Z. Aouissat, G. Chanfray, T. Kunihiro, E. Oset and P. Schuck. A special 
thank goes to M. Vicente-Vacas for the assistance received when using his 
code.
\newpage
%
%
 
\end{document}